# Optical and Electrical Characterization of Boron-Doped Diamond

**Sunil K. Karna[1], Yogesh. K. Vohra[1], and Samuel T. Weir[2]**

[1] Department of Physics, University of Alabama at Birmingham, Birmingham, AL 35294, USA

[2] L-041, Lawrence Livermore National Lab., Livermore, CA 94550, USA

**Abstract**

A homoepitaxial boron-doped single crystal diamond films were grown on synthetic (100) type Ib diamond substrates using a microwave plasma assisted chemical vapor deposition. Raman characterization showed a few additional bands at the lower wavenumber regions along with the zone center optical phonon mode for diamond. The shape modification of the zone center optical phonon line and its downshift were observed with the increasing boron content in the film. A modification in surface morphology of the film with increasing boron content has been observed by atomic force microscopy. Step bunching was appeared on the surface of highly doped film. The difference in activation energy of the carriers indicates two different conduction mechanisms were responsible for the semiconducting behavior of the film.

**Introduction**

Diamond is an excellent research material for extreme conditions due to its chemical inertness and tolerance to extremely harsh conditions of radiation, pressure and temperature. It can change the lifespan and functionality of many future electronic devices if doped with a controlled level of impurity. There have been reports on various levels of doping during the diamond growth process using the microwave plasma assisted chemical vapor deposition (MPCVD) method. Doping can change diamond properties from a wide band gap semiconductor to a metallic conductor and then to a superconductor.[1-5] P-type boron doped polycrystalline diamond has already been demonstrated as a radiation detector, sensor, electrode and electrical heating element in diamond anvil cells.[6-10] However, device performance with polycrystalline diamond is degraded by the presence of grain boundaries and also by the limited thickness of the grown polycrystalline material. To remedy such a problem, the synthesis of high quality single crystal diamond is a widely investigated research field. Hence our study was based on the optimization of growth conditions of semiconducting high quality boron-doped single crystal diamond using MPCVD. In order to understand the effect of boron concentration in the diamond lattice, Raman spectroscopy and electrical conductivity of the films were analyzed in detail.

**Experimental**

Boron doped single crystal diamond films were deposited homoepitaxially on (100) oriented type Ib diamond substrates (misorientation angle less than $1^o$) that were synthesized by high pressure high temperature technique. The diamond seeds of dimensions $3.5 \times 3.5 \times 1.5$ mm$^3$ were used as substrates in a 2.45 GHz bell jar microwave plasma chemical vapor deposition system. The Type Ib substrate (seed crystal) was cleaned with acetone before placing on a shallow cavity of screw shaped molybdenum substrate holder that was mounted on a cooling jacket. The deposition time was 8 hour for all samples. The microwave power was adjusted to 1.5 to 2.5 kW to obtain 1100 to 1200$^o$C substrate deposition temperature. The reactant gas $B_2H_6$ with 6 to 8% of $CH_4$ in $H_2$ mixture was used for the deposition. The $(B/C)_{gas}$ ratio for sample (BD1) was kept at 2300 ppm and that for samples (BD2) and (BD3) was maintained at 6200 ppm. The details of deposition parameters and growth rate of samples (BD1, BD2, and BD3) are summarized in Table 1. Each sample was cleaned in a saturated solution of $CrO_3 + H_2SO_4$ solution at 150$^o$C for 5 minutes, then rinsed in a 1:1 boiling solution of $H_2O_2$ and 30% $NH_4OH$ for 10 minutes, then rinsed again in deionized water and ultrasonicated in acetone for 10 minutes.[11] Finally, the samples were exposed in hydrogen plasma at 900$^o$C for 10 minutes before further characterizations.[12] To avoid any ambiguity due to hydrogenation electrical resistance measurement was taken during cooldown procedure in a four point probe vacuum chamber.

The quality of as deposited films was determined using laser Raman spectroscopy, x-ray rocking curve (omega scan) measurement, and atomic force microscopy (AFM). Raman spectra were recorded using a 514 nm laser excitation wavelength at room temperature. A 5-micron in diameter laser spot was focused on the sample surface to acquire Raman spectrum. For x-ray rocking curve measurements, omega scans were obtained by rotating sample with $0.02^o$ angular step with detector set at a fixed 2theta value corresponding to (400) Bragg diffraction peak from the diamond crystal. Full width at half maximum (FWHM) and shape of rocking curve is a well established quality indicator for homoepitaxial grown diamond. Non-contact mode AFM was used to measure surface topography in this study.

Doping level and electrical conductivity were determined by Fourier transform infrared spectroscopy (FTIR) and four point probe measurement respectively. Four point probe electrical measurements were conducted in a vacuum chamber with a pressure less than 5 mTorr. Nitrogen gas was used to cool the samples in a four probe system. In a four point probe measurement samples were first heated to 690K and reading was taken during cool down process. This procedure was believed to remove the adsorbates from the diamond samples. To measure electrical conductivity, tungsten strips that served as electrical contacts were deposited on the surfaces of the diamond films using a magnetron sputtering technique. Sputtered tungsten adheres well on diamond and its electrical resistivity is about 5 µΩ-cm. An optical microscope image of the tungsten strips on a film surface is shown in Figure 1. In order to minimize self-heating of the samples and ensure ohmic contact conditions with bulk limited currents, the currents through the samples were limited to 0.1 to 1.0 mA.

 Results and discussion

A series of heavily boron doped single crystal diamond were synthesized homoepitaxially with different concentrations and charecterized during this study. The color of the seed crystals was yellowish due to substitutional nitrogen impurity but it changes its color as boron incorporated in its lattice. The color of the boron - doped samples was observed to change from pale blue to dark blue depending upon the boron concentration in the film. AFM images were taken against the flat surfaces of the epitaxial samples over scanned areas of $20 \times 20$ µm$^2$ as shown in Figure 2. The surface morphologies of the doped samples were modified after deposition with polishing scratches from the seed crystals disappearing and grooves types feature beginning to appear. Step bunching is clearly observed on the surface of sample BD3. Early studies show that the step-growth mechanism on (100) face of homoepitaxial diamond may be responsible for the formation of such morphological featueres on diamond film.[13-15]

The Raman spectra of doped and undoped single crystal diamonds are shown in Figure 3. No graphitic carbon related peak was observed in the spectrum indicating high quality homoepitaxial layer. An intense zone-center optical phonon mode of diamond is visible at 1334 cm$^{-1}$ along with additional broad bands at 580, 900, 1042, 1233 cm$^{-1}$. Such bands were previously reported mainly on polycrystalline boron doped diamond.[16,17] A small downshift of bands 1233 and 580 cm$^{-1}$ with increasing boron concentration was also observed in this study. The downshift of optical phonon line and broadening of full width at half maximum (FWHM) with a significant change in intensity of bands 580 and 1233 cm$^{-1}$ were also observed. The origin of those additional bands is not completely understood but the literature shows that bands around 1233 and 580 cm$^{-1}$ are attributed to the presence of large cluster of boron atoms in diamond.[2, 16-18] Gheeraert et al, have described the downshift of optical phonon line with increasing boron content.[16] The asymmetric broadening and downshift of Raman peak of boron – doped diamond films could be explained by Fano-effect.[16,18-21] Fano effect is the quantum mechanical interference between discrete zone-center optical phonon state and continuum of electronic states induced in presence of boron.

The growth rate of diamond was observed to decrease rapidly with the increasing boron content in the film as shown in Table 1. High growth rate up to 23 microns/hour has been observed in undoped sample (HD) in our study. Addition of a little amount of nitrogen and low (B/C)$_{gas}$ ratio in feed gas may be the cause of high growth rate of sample (BD1). On the other hand, the same (B/C)$_{gas}$ ratio in samples (BD2) and (BD3) but less supply of hydrogen in the feed gas of sample and lower substrate temperature reduces the growth rate of sample BD3. Overall, there is a significant decrease in diamond growth rate with an increase in boron content in the plasma. Ramamurti et al, have reported a similar tendency in growth rate of boron doped single crystal diamond.[2]

The quality of the film was also tracked with an x-ray rocking curve experiment as shown in the inset graph of Figure 3. The intense peak of (400) Bragg reflection was observed in the experiment from 30 – 62° omega scans as a manifestation of high quality film. The FWHM of the Bragg peak was also observed to be broadened and it varied from 0.07° to to 0.10° with increasing boron concentration in the film.

The boron concentration was determined from absorption spectrum of FTIR data. Strong absorption was seen in IR Spectra of the samples above 2000 cm$^{-1}$ wavenumber as shown in Figure 4. The one–phonon absorption peak at about 1290 cm$^{-1}$ was used to calculate boron – concentration.[22]

$$[B](cm^{-3}) = (2.1 \times 10^{17}) \times \alpha_{(1290 cm^{-1})} \quad (1)$$

where $\alpha = A/t$, is the linear absorption coefficient at 1290 cm$^{-1}$ band, and $A$ and $t$ being absorbance and thickness of the film. The boron concentrations in film of samples (BD1) and (BD2) was found to be $8.4 \times 10^{18}$ and $3.6 \times 10^{19}$ atom cm$^{-3}$ respectively. Due to strong absorption to infrared signal we were unable to observe the 1290 cm$^{-1}$ absorption band in the FTIR spectrum of sample (BD3); hence its dopant level could not be estimated using equation (1).

The variation in conductivity, (σ) of the films with temperature (T) was determined from four point probe measurement as shown in Figure 5. The room temperature resistivity of the most heavily doped sample (sample (BD3)) was determined to be 0.12 Ωcm. Activation energies of the samples have been obtained by best fit Arrhenius plot of the conductivity data at high and low room temperatures. High and low temperatures are defined here from above and below the transition temperature of the conductivity data as shown in Figure 5. The activation energies of samples (BD1), (BD2), and (BD3) were found to be 0.28 eV, 0.18 eV, and 0.05 eV respectively at high temperature and 0.02, 0.05, and 0.03 eV respectively at low room temperature. The difference in activation energy from high to low temperature indicates that the two different conduction mechanisms are responsible for carrier transport in the film.[23,24] At higher temperatures, carrier is transported via band conduction and at low temperatures, carrier hops in the localized states via hopping conduction.

At high temperatures, the activation energy of carriers was found to decrease with increasing doping concentration. Increasing doping concentration increases the acceptor band width which ultimately reduces the activation energy of acceptors.[18] A transition in the conduction mechanism from localized hopping to band conduction was observed to shift towards higher temperature as the amount of doping increased. Such shift may be the result of a reduction of the carrier mobility in the diamond film with increasing boron concentration. The Pearson and Bardeen formula was also taken into account to estimate the boron concentration, $N_a$ of the films from the calculated activation energy, $E_a$.[25]

$$E_a = E_o - \alpha (N_a)^{1/3} \qquad (2)$$

where $E_o = 0.37 eV$, is the activation energy of an isolated dopant, and $\alpha$ is a material dependent constant whose value for diamond is $6.7 \times 10^{-8}$ eV/cm. The boron concentrations in samples (BD1), (BD2) and (BD3) were estimated using equation (2) to be $2.4 \times 10^{18}$, $2.2 \times 10^{19}$, and $1.0 \times 10^{20}$ atom cm$^{-3}$, respectively. These values lie close to the estimated values of the dopant level determined from the FTIR data at high temperatures.

**Conclusions**

Heavily boron-doped single crystal diamond was synthesized using MPCVD on synthetic Ib diamond seeds with different boron concentrations. Doping concentrations of those samples estimated by the FTIR measurements and from the activation energies measured by the four-probe electrical resistance data are in agreement with each other. The crystalline quality has been confirmed by x-ray rocking curve and Raman scattering. The Raman spectrum of boron-doped diamond films contains additional bands along with first order Raman peak in the lower wavenumber region. The downshift and broadening of Raman lines are also observed with increasing boron concentration in the crystal. The diamond growth rate was observed to decrease with increasing boron content in the film. Temperature dependent resistivity measurements show that the current conduction mechanism depends upon the doping level and obeys semiconductor behavior in the temperature range between 90 to 680 K. A significant difference was found in the activation energies of the films over the given experimental temperature range. A transition in the conduction mechanism from localized hopping to band conduction was observed to shift towards higher temperature as the amount of doping increased. The lowest room temperature electrical resistivity of one of the samples was determined to be 0.12 Ωcm suitable for fabrication of high power electronic devices.

**Acknowledgement:**

This material is based upon work supported by the Department of Energy (DOE) – National Nuclear Security Administration (NNSA) under Grant No. DE-FG52-10NA29660. We are very thankful to Drs. Martyshkin, Stanishevsky and Dashdorje for their help rendered during the characterization of deposited films. Sunil Karna gratefully acknowledged the graduate assistant fellowship from UAB.

**Table Captions:**

**Table 1:** The summary of experimental growth conditions and observed growth rates for various boron-doped diamond films is presented below.

**Figure Captions:**

**Fig. 1.** Optical micrographs of homoepitaxially deposited undoped single crystal diamond sample (HD) and tungsten strips that serve as electrical contacts on doped single crystal diamond sample (BD3).

**Fig. 2**. AFM images of a seed crystal sample (Seed) prior to any CVD growth, homoepitaxially deposited single crystal diamond without boron doping (HD), and boron-doped single crystal diamond samples (BD1), (BD2) and (BD3). The scanned area $20 \times 20$ μm$^2$ in all cases. Surface roughness, $R_{ms}$ of sample (Seed), (HD), (BD1), (BD2) and (BD3) was 3.5, 7.8, 1.2, 4.0, and 1.2 nm respectively.

**Fig. 3**. Raman spectra of seed crystal (Seed), undoped (HD) and doped single crystal diamond films (BD1), (BD2), and (BD3) with different boron concentrations, the position of Raman line and its FWHM is displayed in the graph. Inset is an x-ray rocking curve of sample (BD2).

**Fig. 4**. FTIR spectra of seed crystal (Seed), undoped films (HD), and doped films (BD1), and (BD2). The characteristic boron absorption peak at 1290 cm$^{-1}$ is indicated by a vertical arrow.

**Fig. 5.** Conductivity of samples (BD1), (BD2), and (BD3) as a function of temperatures in the range of 90 to 690K. Arrow sign indicates the transition temperature from the band conduction (high temperatures) to the hopping conduction (low temperatures).

**Table 1**

| Sample | T (°C) | $B_2H_6$ (ppm) | $H_2$ (sccm) | $CH_4$ (sccm) | $O_2$ (sccm) | $N_2$ (sccm) | [r] (μm/hr) |
|---|---|---|---|---|---|---|---|
| (HD) | 1100 | 0 | 400 | 32 | 0.8 | 0.4 | 23 |
| (BD1) | 1100 | 90 | 400 | 32 | 0.8 | 0.4 | 16 |
| (BD2) | 1200 | 190 | 400 | 24 | 0.8 | 0 | 8 |
| (BD3) | 1100 | 190 | 300 | 18 | 0 | 0 | 2 |

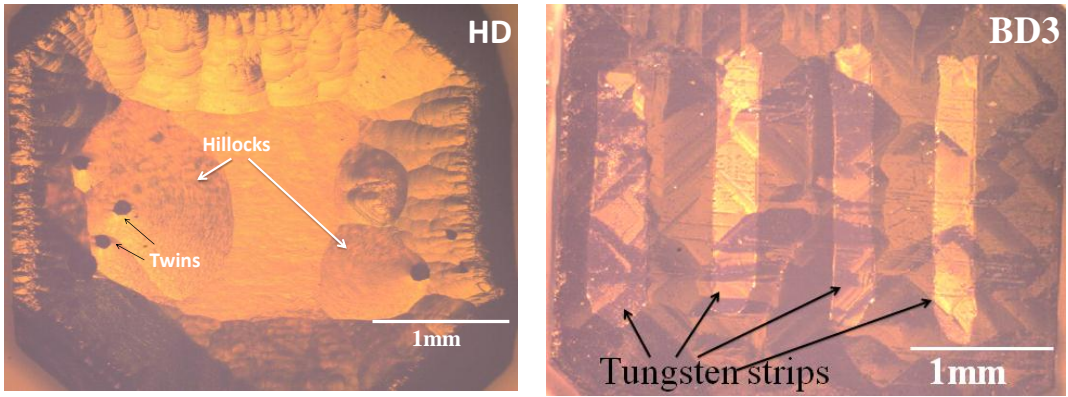

Figure 1

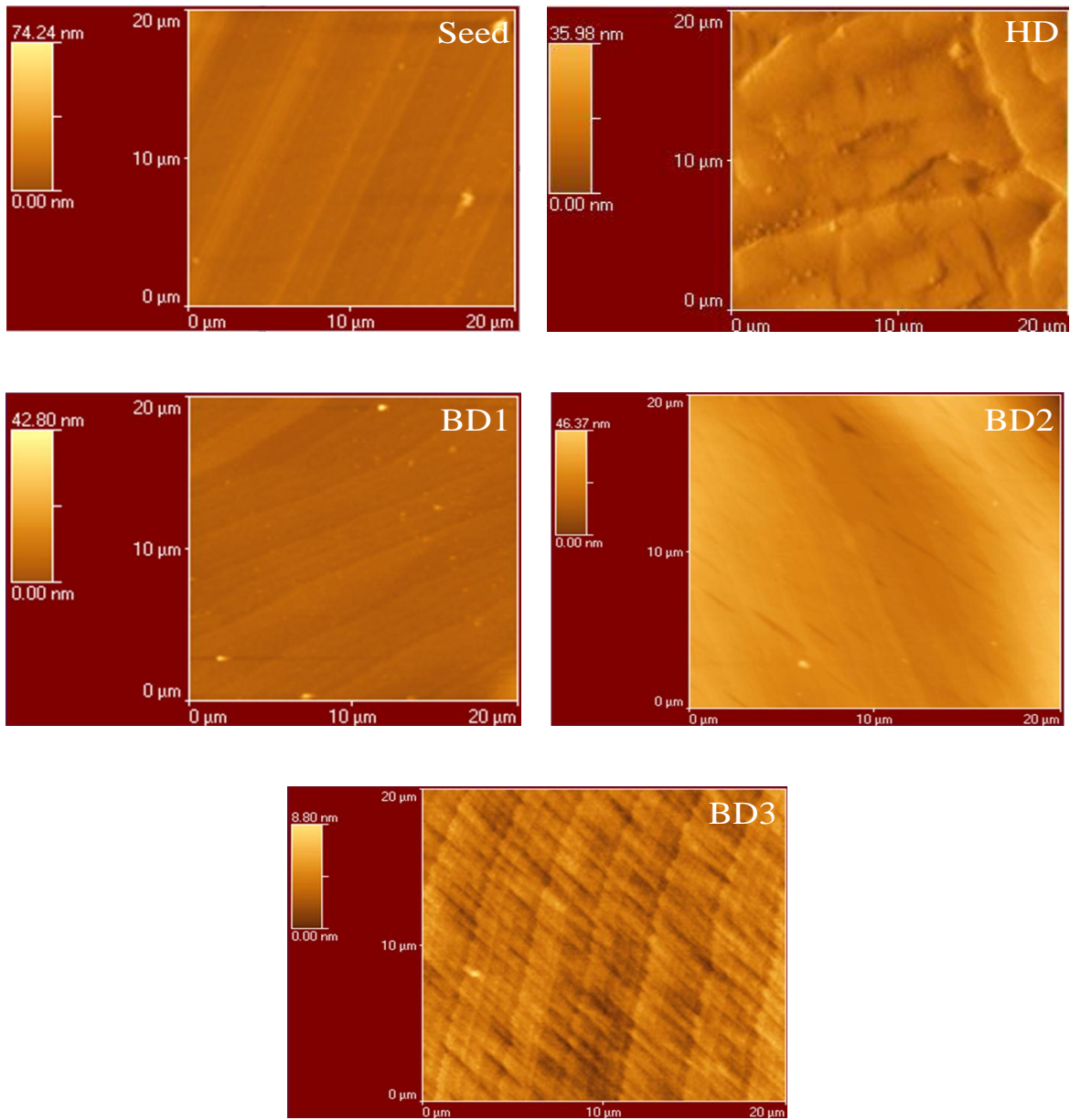

Figure 2

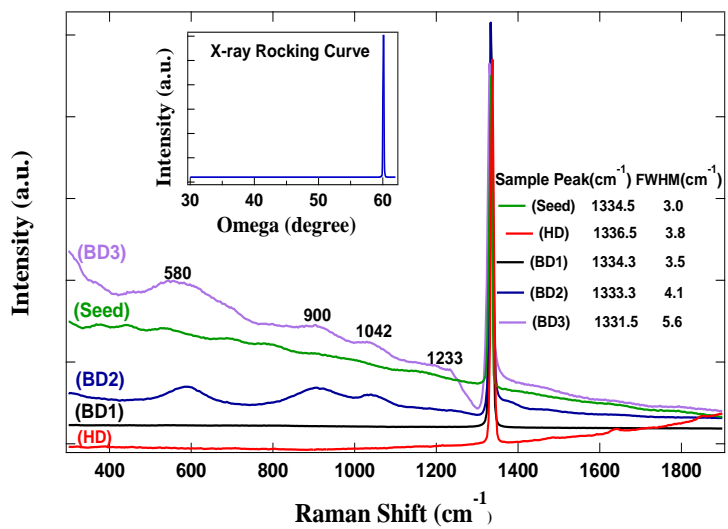

Figure 3

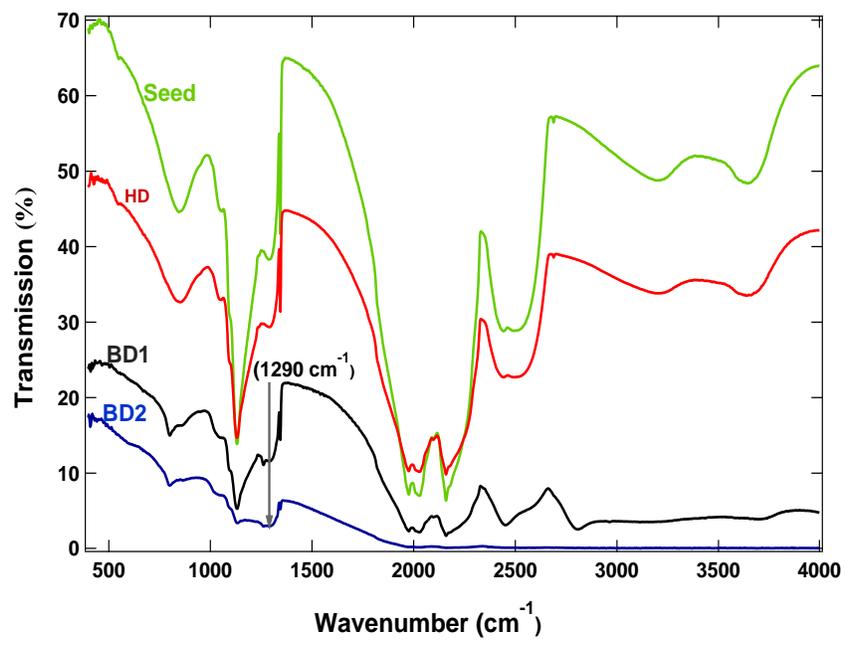

Figure 4

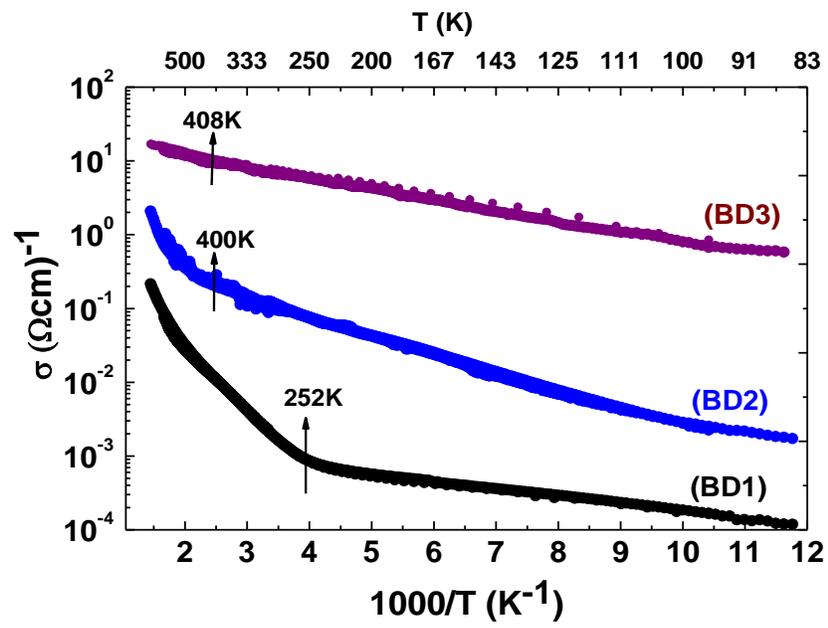

Figure 5